\documentclass[english]{article}
\usepackage[english]{babel}
\usepackage[utf8]{inputenc}
\usepackage{amsmath,amssymb,mathrsfs}
\usepackage{color,array,times}
\usepackage{epsfig,epstopdf,textcomp}
\usepackage{epsfig, subfigure, color}
\usepackage{float, multirow}
\usepackage{amsmath}
\usepackage{enumitem,kantlipsum}
\usepackage{graphicx}
\usepackage{epsfig, subfigure, color}
\usepackage{float, multirow}
\usepackage{epstopdf}
\DeclareGraphicsExtensions{.eps}
\usepackage{amssymb}
\usepackage{algorithm,algorithmic}
\usepackage{tikz}
\usetikzlibrary{matrix}
\usepackage{cancel}
\usepackage{systeme}
\usepackage{amsmath,stackengine}
\stackMath

\begin{document}

\title{Decoding of Polar Codes Based on Q-Learning-Driven Belief Propagation}

\author{Lucas M. de Oliveira, Robert M Oliveira and Rodrigo C. de Lamare
\thanks{Lucas M. de Oliveira, Robert M Oliveira and Rodrigo C. de Lamare - Centre for Telecommunications Studies (CETUC), Pontifical Catholic University of Rio de Janeiro (PUC-Rio), Rio de Janeiro-RJ, Brasil, E-mails: lucasmarques@cetuc.puc-rio.br, rbtmota@gmail.com and delamare@cetuc.puc-rio.br. This work was supported by CNPq, CAPES, FAPERJ and FAPESP.} }

\maketitle



\begin{abstract}
This paper presents an enhanced belief propagation (BP) decoding algorithm and a reinforcement learning-based BP decoding algorithm for polar codes. The enhanced BP algorithm weighs each Processing Element (PE) input based on their signals and Euclidean distances using a heuristic metric. The proposed reinforcement learning-based BP decoding strategy relies on reweighting the messages and consists of two steps: we first weight each PE input based on their signals and Euclidean distances using a heuristic metric, then a Q-learning algorithm (QLBP) is employed to figure out the best correction factor for successful decoding. Simulations show that the proposed enhanced BP and QLBP decoders outperform the successive cancellation (SC) and belief propagation (BP) decoders, and approach the SCL decoders.
\end{abstract}


\section{Introdution}

Polar codes, originally introduced in 2009 by Arikan \cite{ref1}, are a significant breakthrough in coding theory. They are theoretically proven capacity-achieving codes based on the general channel polarization phenomenon \cite{ref1}. As part of the 5G New Radio enhanced mobile broadband (eMBB) standard, significant research efforts have been made to design satisfactory decoders to meet low-latency and high-speed requirements, ranging from efficient decoding to suitable hardware implementation.

One of the first decoders that arose was the Successive Cancellation (SC) decoder \cite{ref1}, which can achieve good error-correcting capability with low complexity. However, due to the type of SC-based decoding characterized by serial message updating, propagation errors and low capacity for high-speed real-time applications this decoder often exhibits low performance. Therefore, the successive cancellation list (SCL) decoding \cite{ref2} was proposed to improve the error-correction performance of SC, since it stores the most likely codewords in a list, reducing error probability and improving the performance. Moreover, SCL can be further enhanced by concatenating a cyclic redundancy check (CRC) code \cite{ref2}. As can be seen in \cite{ref3} - \cite{ref5} , CRC-aided successive cancellation list (CA-SCL) decoding attains promising error-correction performance. 

Furthermore, several attempts have been made to reduce the computational complexity and increase the throughput of SC and SCL decoders. Inherited from Low-Density Parity-Check (LDPC) codes, Belief Propagation (BP) decoders were introduced in \cite{ref6}, because of their particular advantages with respect to parallelism, high throughput, and low latency. Nevertheless, due to their characteristics, BP decoding requires a large number of iterations to achieve good performance. Thus, a way to improve the performance is to employ BP list decoding \cite{ref7}, which operates when the standard polar code factor graph fails to produce the correct decoding result and the permuted version of the standard graph may yield the correct estimate.

\par In this paper, we propose an enhanced BP algorithm and a Q-Learning BP (QLBP) approach to enhance BP decoding of polar codes. Initially, a weighting technique based on the Euclidean distance and the signal of the Processing Element inputs is presented and incorporated into a BP strategy to devise the enhanced BP algorithm. Then, based on the fact that a correction factor can enhance the weighting process, the QLBP is devised to compute the best factor and to ensure an optimized decoding performance. Numerical results show that the proposed QLBP algorithm outperforms the proposed enhanced BP, the existing BP and the SC decoding algorithms.

The remainder of this paper is organized as follows. Section II introduces polar codes and the decoding problem with BP. Section III presents the Enhanced BP algorithm and its weighting method. Section IV proposes a Q-learning strategy for computing BP weights, then presents the Q-Learning driven BP decoding algorithm. Section V presents the simulated results. Conclusions are drawn in Section VI.

\section{Preliminaries}

\subsection{Polar Codes}

Polar codes are derived from channel combination and polarization theory. As the code length N=$2^n$ gets larger through splitting and combining channels, the symmetric capacity of bit-channels tends to either 1 or 0. In that way, there are basically two types of channels: noiseless channels, closer to the capacity of the binary symmetric channels and denoted by the set $\mathbb{A}$, and noisy channels, denoted by the set $\overline{\mathbb{A}}$. Let ${\mathbf{u}}_1^n=\{u_1,u_2,...,u_n\}$ denote the source vector and ${\mathbf{x}}_1^n=\{x_1,x_2,...,x_n\}$ denote the code word vector. For polar codes with $\mathcal{P}$(N,K), R=$\frac{\text{K}}{\text{N}}$ ,the vector ${\bf u}$ consists of K information bits in $\mathbb{A}$ and N-K frozen bits in $\overline{\mathbb{A}}$. The encoding process of polar codes, defined by Arikan, can be expressed by $\mathbf{x}_1^n=\mathbf{u}_1^n {\mathbf{G}}^{\otimes n}$, where $\mathbf{G}= \begin{pmatrix}
1 & 0 \\
1 & 1
\end{pmatrix}$ is the n-th Kronecker power of the polarizing matrix $\mathbf{G}$ and $\mathbf{n}=\log_2 \text{N}$. 

\subsection{Belief Propagation Decoding}
\label{bpdecoding}

The BP decoder is a message-passing decoder with iterative processing over the factor graph of any polar code $\mathcal{P}$(N,K) that has found numerous applications in wireless communications \cite{bfpeg,rrser,rootldpc,memd,baplnc,dopeg,jidf,spa,mbdf,mbthp,bfidd,vfap,kaids,1bitidd,did,lrcc,aaidd,listmtc,dynovs,rcpd,detmtc,srbars,dynmtc,nupd}. The factor graph is based on corresponding polarization matrix $\mathbf{G}^{\otimes n}$, composed of $n =\log_2 N$ stages, each one with $N/2$ processing elements (PEs),  and $(n+1)N$ nodes. Two types of LLRs are transmitted over the factor graph: the left-to-right message $R_{i,j}^{(t)}$ and the right-to-left message $L_{i,j}^{(t)}$, where $i,j$ denotes the j-th node at the i-th stage whereas $t$ denotes the t-th iteration.

\par Considering a binary phase-shift keying(BPSK) modulation and additive white Gaussian noise (AWGN) channel model, the noisy received code word is given by
\begin{equation}
    \textbf{y}=(\textbf{1}-2\textbf{x})+\textbf{z}
\end{equation}
where \textbf{1} is an all-one vector, \textbf{z} is the AWGN noise vector with variance $\sigma^2$ and zero mean. In LLR domain, the LLR inputs for BP decoding of polar codes are initialized as: 

\begin{subequations}
	 \renewcommand{\theequation}{\theparentequation.\arabic{equation}}
	\label{eq2} 
    \begin{align}
	   &L_{n+1,j}^{(0)}=\ln \frac{P_r(x_j=0|y_j)}{P_r(x_j=1|y_j)}=\frac{2y_j}{\sigma^2}\\
       &R_{1,j}^{(0)}=\systeme{0 \ \ \ \text{if j} \in \mathbb{A}, +\infty \ \ \ \text{if j} \in \overline{\mathbb{A}}}
	 \end{align}
\end{subequations}
where $x_j$ and $y_j$ denote the j-th bit of modulated and received codeword, respectively.

\begin{figure}[H]
		    \centering
	        \includegraphics[width=2.5in]{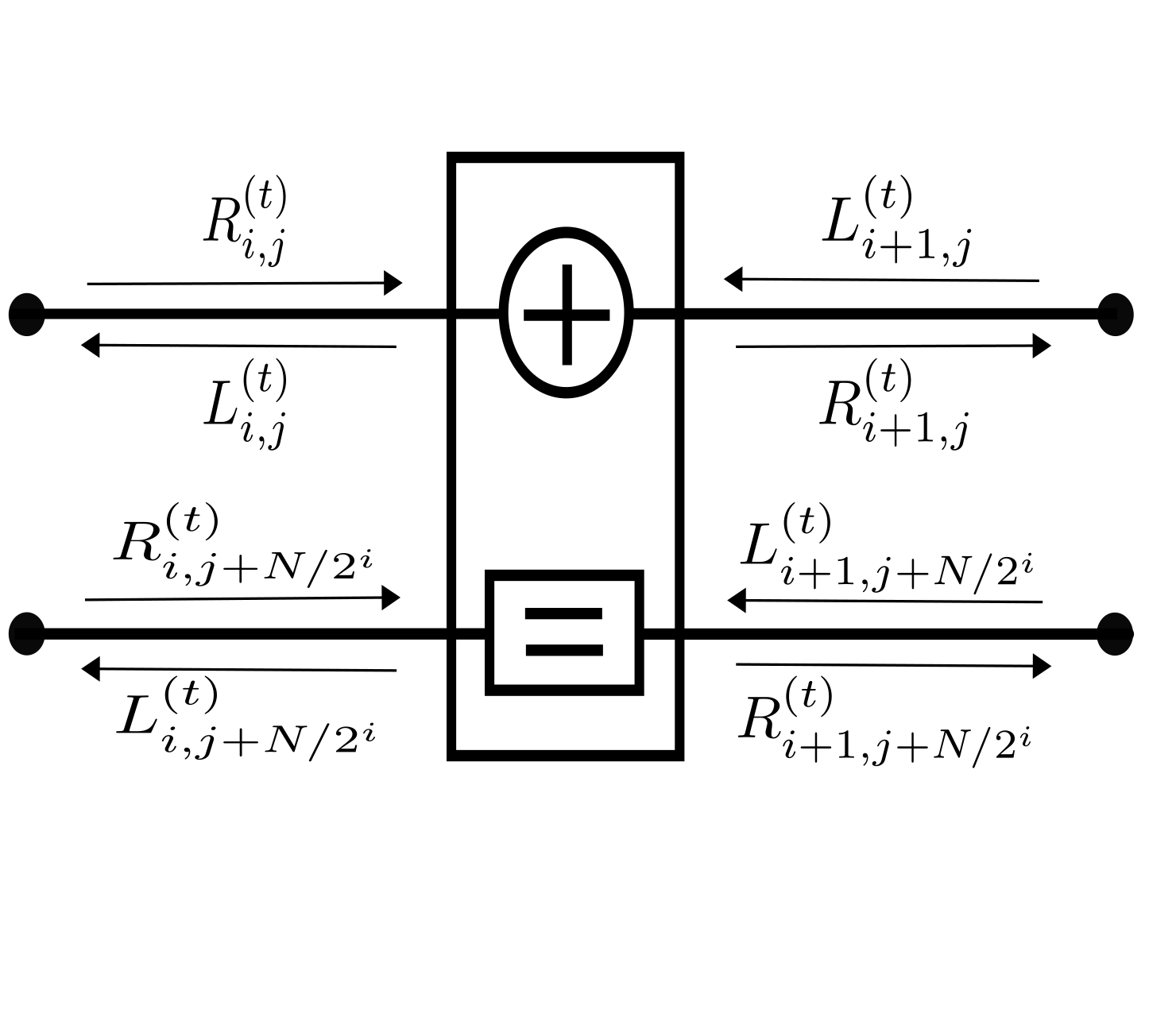} 
	       \vspace{-3em} \caption{Processing element update} 
	        \label{figura1} 
        \end{figure}

The forward and backward propagation of the LLRs over the PEs, shown in figure \ref{figura1}, is based on the following iterative updating rules:

\begin{equation}
	 \begin{aligned}
\label{eq3}
	   &L_{i,j}^{(t)}=g(L_{i+1,j}^{(t-1)},L_{i+1,j+N/2^i}^{(t-1)}+R_{i,j+N/2^i}^{(t-1)})\\
       &L_{i,j+N/2^i}^{(t)}=g(L_{i+1,j}^{(t-1)},R_{i,j}^{(t-1)})+L_{i+1,j+N/2^i}^{(t-1)}\\
       &R_{i+1,j}^{(t)}=g(R_{i,j}^{(t-1)},L_{i+1,j+N/2^i}^{(t-1)}+R_{i,j+N/2^i}^{(t-1)})\\
       &R_{i+1,j+N/2^i}^{(t)}=g(L_{i+1,j}^{(t-1)}+R_{i,j}^{(t-1)})+R_{i,j+N/2^i}^{(t-1)}
	 \end{aligned}
\end{equation}
where $g(x,y)$ is referred to as the operator:
\begin{equation}
\label{eq4}
\begin{aligned}
    g(x,y)&=\ln \frac{1+e^{x+y}}{e^x+e^y} \\
    &\approx 0.9375\cdot sign(x)\cdot sign(y)\cdot \min (|x|,|y|)
    \end{aligned}
\end{equation}

\vspace{-0.35in}
        
When the maximum number of iterations, $T_{max}$, is reached, the information bit ${\mathbf{\hat{u}}}_j$ and the transmitted codeword ${\mathbf{\hat{x}}}_j$ are estimated based on their LLRs using the following hard decision criteria:
\begin{subequations}
	 \renewcommand{\theequation}{\theparentequation.\arabic{equation}}
\begin{align}
    \label{eq51}
    &\hat{u}_j=\begin{cases} 
    0, & \mbox{if } L_{1,j}^{T_{max}}+R_{1,j}^{T_{max}} > 0 \\ 1, & \mbox{otherwise }
    \end{cases}
    \\
    &\hat{x}_j=\begin{cases} 
    0, & \mbox{if } L_{n+1,j}^{T_{max}}+R_{n+1,j}^{T_{max}} > 0 \\ 1, & \mbox{otherwise }
    \end{cases}
\end{align}
\end{subequations}

\section{Enhanced BP Decoding} 
\label{Enhanced BP Decoding}

In this section, we propose a weighting technique for BP decoders. This technique will lay the foundation for the Q-learning algorithm, which will be discussed later. As can be seen in equation \eqref{eq3}, the propagation of the messages requires four parameters for each direction, which is, $L_{i+1,j}^{(t)}$, $L_{i+1,j+N/2^i}^{(t)}$, $R_{i,j}^{(t)}$ and $R_{i,j+N/2^i}^{(t)}$ for L messages and $R_{i,j}^{(t)}$, $R_{i,j+N/2^i}^{(t)}$, $L_{i+1,j}^{(t)}$ and $L_{i+1,j+N/2^i}^{(t)}$ for R messages. Moreover, as the PEs updates can be summarized in signal and modules successive operations defined by equation \eqref{eq3}, our proposed weighting technique is built on how those four LLRs evolve in terms of signal and module over time.   

     Thus, we introduce four weighting factors, $\rho_1, \rho_2, \rho_3$, and $\rho_4$, which will modify how the LLRs are updated, as can be seen below.

    \begin{equation}
	 \begin{aligned}
    \label{eq6}
	   &L_{i,j}^{(t)}=g\left(\rho_1\cdot L_{i+1,j}^{(t-1)},\rho_1 \left(L_{i+1,j+N/2^i}^{(t-1)}+R_{i,j+N/2^i}^{(t)}\right)\right)\\
       &L_{i,j+N/2^i}^{(t)}=g\left(\rho_2 \cdot L_{i+1,j}^{(t-1)},\rho_2\cdot R_{i,j}^{(t)}\right)+\rho_2 \cdot L_{i+1,j+N/2^i}^{(t-1)}
	 \end{aligned}
\end{equation}

    where 
    \begin{equation}
\label{eq7}
\begin{aligned}
&\rho_1=1+\beta\cdot \left[\frac{||L_{i,j}^{(t)}|-|L_{i,j}^{(t-1)}||}{(|L_{i,j}^{(t)}|+|L_{i,j}^{(t-1)}|)}\right]\cdot \Delta_1\\ 
&\rho_2=1+\beta\cdot \left[\frac{||L_{i,j+N/2^i}^{(t)}|-|L_{i,j+N/2^i}^{(t-1)}||}{(|L_{i,j+N/2^i}^{(t)}|+|L_{i,j+N/2^i}^{(t-1)}|)}\right]\cdot \Delta_2\\
&\Delta_1=sign(L_{i,j}^{(t)}+L_{i,j}^{(t-1)})\\
&\Delta_2=sign(L_{i,j+N/2^i}^{(t)}+L_{i,j+N/2^i}^{(t-1)})
\end{aligned}
\end{equation}

     The weighting method is based on the distance between the LLRs at the time $t$ and $t-1$ and whether their signals have changed over the iterations. Note that when $|L_{i,j}^{(t)}|$ and $|L_{i,j}^{(t-1)}|$ are close to each other, $\rho_1$ is approximately equal to 1. Consequently, the node update is similar to the equation \ref{eq3}. Thus, as these values deviate, the greater the weighting. Moreover, the signal deviations are considered by $\Delta_1$. It should also be pointed out that the same idea is applied to the subsequent weighting factors.

    \begin{equation}
        \begin{aligned}
        \label{eq8}
        &R_{i+1,j}^{(t)}=g\left(\rho_3 \cdot R_{i,j}^{(t)},\rho_3\left(L_{i+1,j+N/2^i}^{(t-1)}+R_{i,j+N/2^i}^{(t)}\right)\right)\\
        &R_{i+1,j+N/2^i}^{(t)}=g\left(\rho_4\left(L_{i+1,j}^{(t-1)}+R_{i,j}^{(t)}\right)\right)+\rho_4 \cdot R_{i,j+N/2^i}^{(t)}
        \end{aligned}
    \end{equation}
    
    where
    \begin{equation}
    \label{eq9}
        \begin{aligned}
            &\rho_3=1+\beta\cdot \left[\frac{||R_{i+1,j}^{(t)}|-|R_{i+1,j}^{(t-1)}||}{(|R_{i+1,j}^{(t)}|+|R_{i+1,j}^{(t-1)}|)}\right] \cdot \Delta_3\\
            &\rho_4=1+\beta\cdot \left[\frac{||R_{i+1,j+N/2^i}^{(t)}|-|R_{i+1,j+N/2^i}^{(t-1)}||}{(|R_{i+1,j+N/2^i}^{(t)}|+|R_{i+1,j+N/2^i}^{(t-1)}|)}\right]\cdot \Delta_4\\
            &\Delta_3=sign(R_{i+1,j}^{(t)}+R_{i+1,j}^{(t-1)})\\
            &\Delta_4=sign(R_{i+1,j+N/2^i}^{(t)}+R_{i+1,j+N/2^i}^{(t-1)})
        \end{aligned}
    \end{equation}

     Furthermore, it is worth noting that $\beta$ is a general correction factor for all processing elements, whose simulations have shown that it must belong to the range [-0.50, 0.50]. Thus, an open problem is how to set up the best $\beta$ for a specific input, a task in which the Q-learning algorithm, proposed in the next section, tries to solve. 

    A high-level description of the enhanced BP decoding algorithm is illustrated in Algorithm 1. The algorithm takes the received codeword ${\mathbf{y}}_1^n$, the code block length N, the maximum number of iterations $T_{max}$, and the information set $\mathbb{A}$ and calculates the estimated free bits ${\mathbf{\hat{u}}}_\mathbb{A}$ as an output vector.

    \begin{algorithm}[h]
     \caption{Enhanced BP Algorithm}
     \begin{algorithmic}[1]
     \renewcommand{\algorithmicrequire}{\textbf{Input:}}
     \renewcommand{\algorithmicensure}{\textbf{Output:}}
     \REQUIRE ${\mathbf{y}}_1^n$, N, $T_{max}$, $\mathbb{A}$
     \ENSURE  ${\mathbf{\hat{u}}}_\mathbb{A}$
        \FOR {each node (i,j)}
            \IF{(i==1) \& (j $\not \in \mathbb{A}$)}
                \STATE $R_{1,j}^{(0)} \leftarrow \infty$
            \ELSIF {(i==n+1)}
                \STATE $L_{n+1,j}^{(0)} \leftarrow \frac{2y_j}{\sigma^2}$
            \ELSE
                \STATE $L_{i,j}^{(0)} \leftarrow 0$ , $R_{i,j}^{(0)} \leftarrow 0$
            \ENDIF
        \ENDFOR
        \FOR {(1$<$t$<T_{max}$) \& each node (i,j)}
            \IF {(t==1)}
                \STATE Update $L_{i,j}^{(1)}$ and $R_{i,j}^{(1)}$ according to equation $\eqref{eq3}$
                \STATE Store $L_{i,j}^{(1)}$ and $R_{i,j}^{(1)}$
            \ELSE
                \STATE Update $L_{i,j}^{(t)}$ and $R_{i,j}^{(t)}$ according to equations $\eqref{eq6}$ and $\eqref{eq8}$, respectively
                \STATE Store $L_{i,j}^{(t)}$ and $R_{i,j}^{(t)}$
            \ENDIF
        \ENDFOR
        \STATE Compute ${\mathbf{\hat{u}}}$ according to equation \eqref{eq51} 
        \STATE Select $\mathbb{A}$ positions of ${\mathbf{\hat{u}}}$ to compose ${\mathbf{\hat{u}}}_\mathbb{A}$
     \RETURN{} ${\mathbf{\hat{u}}}_\mathbb{A}$
     \end{algorithmic} 
     \end{algorithm}

\section{Proposed Q-Learning BP Decoding}

\subsection{Reinforcement learning and Q-Learning} 
    \label{RLandQ-Learning}

    Reinforcement Learning, RL, is an area of machine learning in which an agent learns how to take actions from its action space, within a particular environment, in order to maximize rewards over time. At each time step, the agent, which is undergoing the learning process,  is in a state $s_t$, selects an action $a_t$ and moves to the next state $s_{t+1}$, while obtaining a reward $r_t$. The aim of learning is to train the agent to find an optimal policy, which is a mapping between states and actions, and will return the maximum cumulative rewards from taking a series of actions in one or more states.

    A Markov Decision Process \cite{ref8}, MDP, is a mathematical framework for fully observable sequential decision making problems in stochastic environments. Defined as a 5-tuple, $(S$,$A$,$R$,$P(s,a,s')$,$R(s,a,s'))$, $S$ represents a set of states, where $s_t \in S$ is the state at time-step $t$, $A$ is a finite set of actions, where $a_t \in A$ is the action executed at time-step $t$, $P(s,a,s')$ is the probability that action $a$ in state $s$ at time $t$ will lead to state $s'$ at time $t+1$, and $R(s,a,s')$ is the immediate reward received after a transition from state $s$ to $s'$ due to action $a$.

     Q-Learning \cite{ref9}, a model-free reinforcement learning algorithm, is used to learning the optimal policy of an agent without using or estimating the dynamics of the environment. For every state-action pair a Q-value, Q($s$,$a$), measures the total amount of discounted rewards expected over the future when the agent moves from the state $s_t$ to the state $s_{t+1}$ with $a_t$ and sticks to its policy afterwards. The Q-learning update rule assumes the following general form:  
    \begin{equation}
        \begin{aligned}
        Q^n(s_t,a_t) &= Q^o(s_t,a_t)+\alpha \delta \\
        \delta &= \left[r_t+\gamma \cdot \max_{a} Q(s_{t+1},a)-Q(s_t,a_t)\right]
        \label{qupdate}
        \end{aligned}
    \end{equation}
    
    where $\alpha$ is the learning rate, and affects how the Q-values are altered after taking an action. The constant $\gamma$ is the discount factor, and determines how much influence the future rewards have on the updates of the Q-values.  

\subsection{Q-Learning BP Decoding}

As mentioned in section \ref{Enhanced BP Decoding}, an open problem is how to compute the best value of $\beta$. However, instead of having a general value of $\beta$ to all PEs on the factor graph, we evaluated different values of $\beta$ for each PE in order to avoid a huge state subspace. From this approach, our environment is composed by each $\text{PE}_{l,m}$, $l\text{, }m$ denotes m-th PE at the l-stage, where $m$ belongs to the range $[1, \frac{N}{2}]$ and $l$ belongs to the range $[1,n]$. Thus, the agent, in a given state $s_{t_{l,m}}$, selects an action $a_{t_{l,m}}$ and waits for the decoding process to return to $\text{PE}_{l,m}$ to be ultimately rewarded.

\begin{itemize}[wide, labelwidth=!, labelindent=0pt]
    \vspace{0.1in}
    \item \textbf{Reward}:
    
     The reward quantifies the desirability of choosing
an action while transitioning to some state. It can be either positive or negative, the latter being interpreted as a penalty for an undesirable action. The total reward with the discount factor that the agent will achieve from the current time step t to the end of the task can be defined as:
\begin{equation}
    R_t=r_t+\gamma r_{t+1}+...+\gamma^{n-t} r_n=r_t+\gamma R_{t+1}
\end{equation}
    
     We have implemented a high positive reward for a successful decoding process in order to encourage the agent to achieve this goal. In addition, the agent should obtain a slight positive reward if the LLR at time $t$ has the same signal of the LLR at time $t-1$ in a given PE. On the other hand, if the LLR at time $t$ and $t-1$ has not the same signal, the agent should obtain a slight negative reward. In doing so, we avoid undesirable actions, which keep away from successful decoding. The discount factor and the rewards values are shown in Table I.

   \begin{table}[ht]
\caption{} 
\centering
\begin{tabular}{c c} 
\hline\hline 
\textbf{Event} & \textbf{Reward} \\ [0.5ex] 
\hline 
Successful decoding & 20/10/0 \\ 
LLRs have the same signal & 1 \\
LLRs have not the same signal & -1 \\
Discount factor & 0.6 \\ [1ex] 
\hline
\end{tabular}
\label{table:nonlin} 
\end{table}

  Note that when we have successful decoding, the reward can be either 20, 10, or 0. As can be seen in section \ref{bpdecoding}, the propagation of the LLRs over the PEs involves four numbers and generates two output LLRs for each propagation direction. Therefore, if both output LLRs have not changed their signal over time, we reward the associated PE with 20. If one of them has changed its signal over time, we reward the associated PE with 10. On the other hand, if both output LLRs have changed their signal, we reward with 0.  
  
    \vspace{0.1in}
    \item \textbf{State Space}:
    
    The state space is the set of all possible situations a processing element could have. For each propagation, it is necessary four numbers to compute the outputs. Thus, there are $2^4$ signal variations and $4!$ modules variations, which means a state space with $4!\cdot 2^4$= 384 possible states.
    
    \vspace{0.1in}
    \item \textbf{Actions}:

    Note that the agent cannot control so far what state it ends up in, since it can be influenced by choosing some action $a$. Thus, focusing on a single state $s$ and action $a$, we introduce recursively the Q function, $Q(s,a)$, in terms of the Q-value of the next state $s'$, which can be expressed as follows:  

    \begin{equation}
        Q(s,a)=r+\gamma \cdot \text{max}_{a'}Q(s',a')
        \label{bellman}
    \end{equation}
    
     Also known as the Bellman equation, equation \eqref{bellman} tells us that the maximum future reward is given by the reward the agent received for entering the current state $s$ plus the maximum future reward for the next state $s^{'}$. 

     In the proposed QLBP algorithm, during the decoding process, for each PE in the factor graph, the agent encounters one of the 384 states and it takes an action. The action in our case can be a value between -0.5 to 0.5. If the signal of the LLR at time $t$ is different from the LLR at $t-1$, the choice of an action will change how the LLR is weighted according to equations \eqref{eq7} and \eqref{eq9}. Otherwise, it is assumed $\rho_{1,2,3,4}$ = 1.

     An agent could interact with the environment in 2 different ways. The first one is to use a lookup table with state-action pairs to store and get information. Each state-action pair is associated with a Q-value that indicates the quality of the decision. Thus, in a given state $s^{*}$, the agent selects the action based on the maximum value of Q($s^{*},a$). This procedure is known as exploiting since we use the information we have available to us to make a decision.

     The second way to take action is to act randomly. This is called exploring and the exploration method used is the greedy approach \cite{ref10}. Instead of selecting actions based on the maximum future reward, we select an action at random. Acting randomly is important because it allows the agent to explore and discover new states that otherwise may not be selected during the exploitation process. It is possible to balance exploration and exploitation using $\epsilon$, which measures how often you want to explore instead of exploit. It was used $\epsilon=0.5$.

\end{itemize}

 A high-level description of the QLBP algorithm is depicted in Algorithm 2. The algorithm takes the conventional BP parameters (such as the received codeword ${\mathbf{y}}_1^n$, the code block length N, the maximum number of iterations $T_{max}$, and the information set $\mathbb{A}$), the Q-Table and the Q-Learning parameters (such as learning rate $\alpha$, discount factor $\gamma$, and $\epsilon$-greedy parameter), and outputs the estimated free bits ${\mathbf{\hat{u}}}_\mathbb{A}$ and the actualized Q-Table. 

\begin{algorithm}[h]
 \caption{QLBP Algorithm}
 \begin{algorithmic}[2]
 \renewcommand{\algorithmicrequire}{\textbf{Input:}}
 \renewcommand{\algorithmicensure}{\textbf{Output:}}
 \REQUIRE Conventional BP parameters, Q-Table, Q-Learning parameters
 \ENSURE  ${\mathbf{\hat{u}}}_\mathbb{A}$, Q-Table
  \STATE Define conventional BP initialization
        \FOR {(1$<$t$<T_{max}$)}
            \FOR {each node $(i,j)$}
                \STATE Update $L_{i,j}^{(t)}$ and $R_{i,j}^{(t)}$ according to equation $\eqref{eq6}$ and $\eqref{eq8}$, respectively
                \STATE Store $L_{i,j}^{(t)}$ and $R_{i,j}^{(t)}$
    
                \IF {rand()$<\epsilon$} 
                    \STATE Choose an action randomly
                \ELSE
                    \STATE Choose the action which maximizes the Q-value for the current state
                \ENDIF
                \IF {$sign(LLR_{i,j}^{(t)}) \neq  sign(LLR_{i,j}^{(t-1)})$}
                    \STATE Apply a penalty
                \ELSE
                    \STATE Apply a reward
                \ENDIF
                \STATE $Q^{n}(s_t,a_t)$=$Q^{o}(s_t,a_t)$+$\alpha(r_t$+$\gamma$ $\cdot \max_aQ(s_{t+1},a_t))$
            \ENDFOR
            \IF {($x=u \cdot G$)}
              	\STATE Apply a bigger reward
              	\STATE Compute $Q^{n}(s_t,a_t)$ again
              	\STATE break
            \ENDIF
        \ENDFOR
        \STATE Compute ${\mathbf{\hat{u}}}$ according to equation \eqref{eq51} 
        \STATE Select $\mathbb{A}$ positions of ${\mathbf{\hat{u}}}$ to compose ${\mathbf{\hat{u}}}_\mathbb{A}$
 \RETURN [${\mathbf{\hat{u}}}_\mathbb{A}$,actualized Q-Table]
 \end{algorithmic} 
 \end{algorithm}

\section{Simulations}

    \hspace{0.2in} In this section, the simulation results are presented to demonstrate the effectiveness of the proposed enhanced BP and QLBP algorithms compared to different decoders, namely Arikan's original SC, SCL with List-4 and List-8 and Arikan's original BP for N=256, N=512 and R=$\frac{1}{2}$. 
    
\begin{figure}[h!]
		    \centering
	        \includegraphics[width=2.5in]{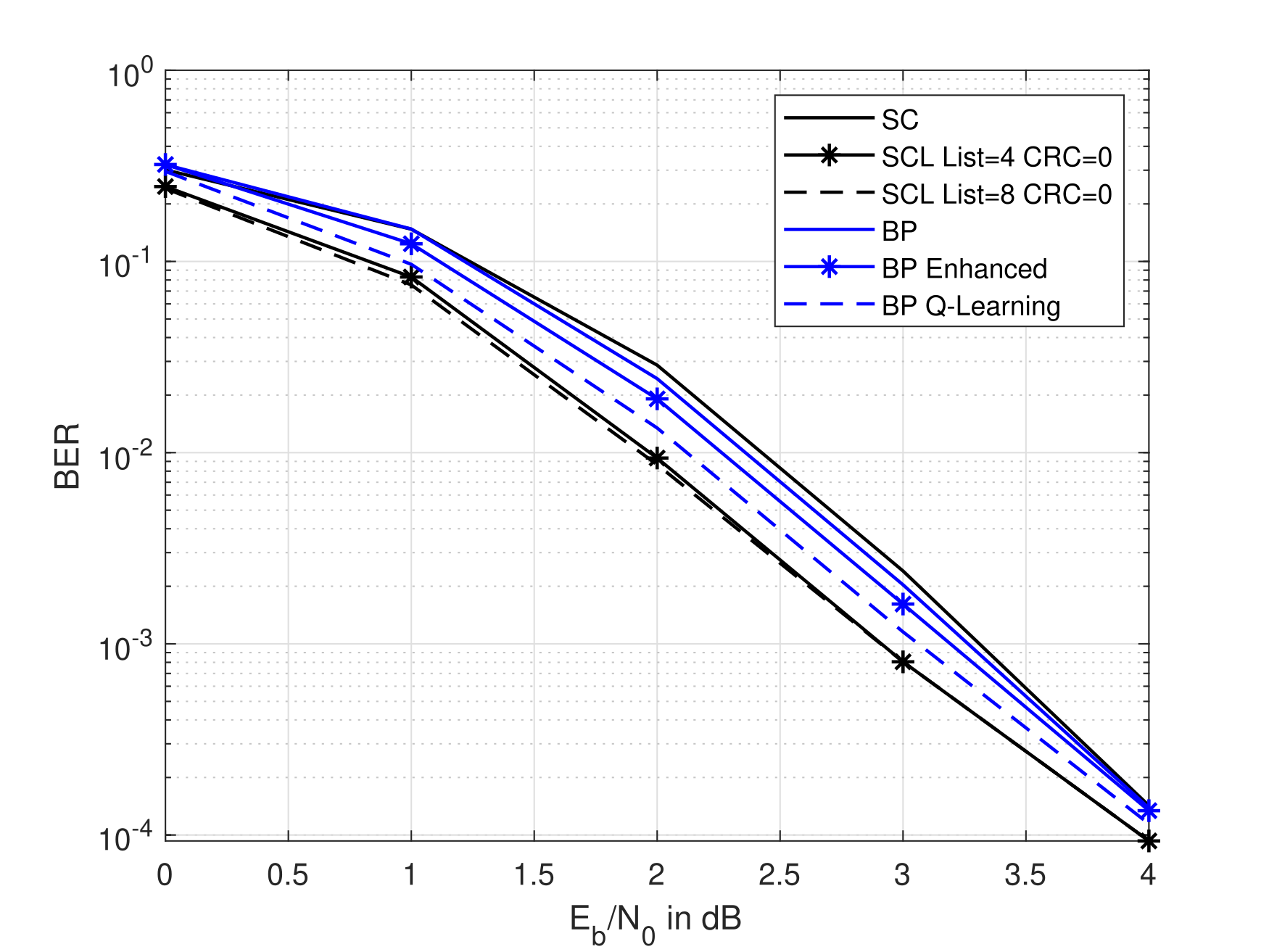} 
	        \caption{BER comparison between Arikan's original SC, SCL with List-4 and List-8, Arikan's original BP,  our proposed BP Enhanced and BP Q-learning decoders for N=256 and K=128; no CRC used.} 
	        \label{figura2} 
	        \vspace{-0.10in}
        \end{figure}

\begin{figure}[h!]
		    \centering
	        \includegraphics[width=2.5in]{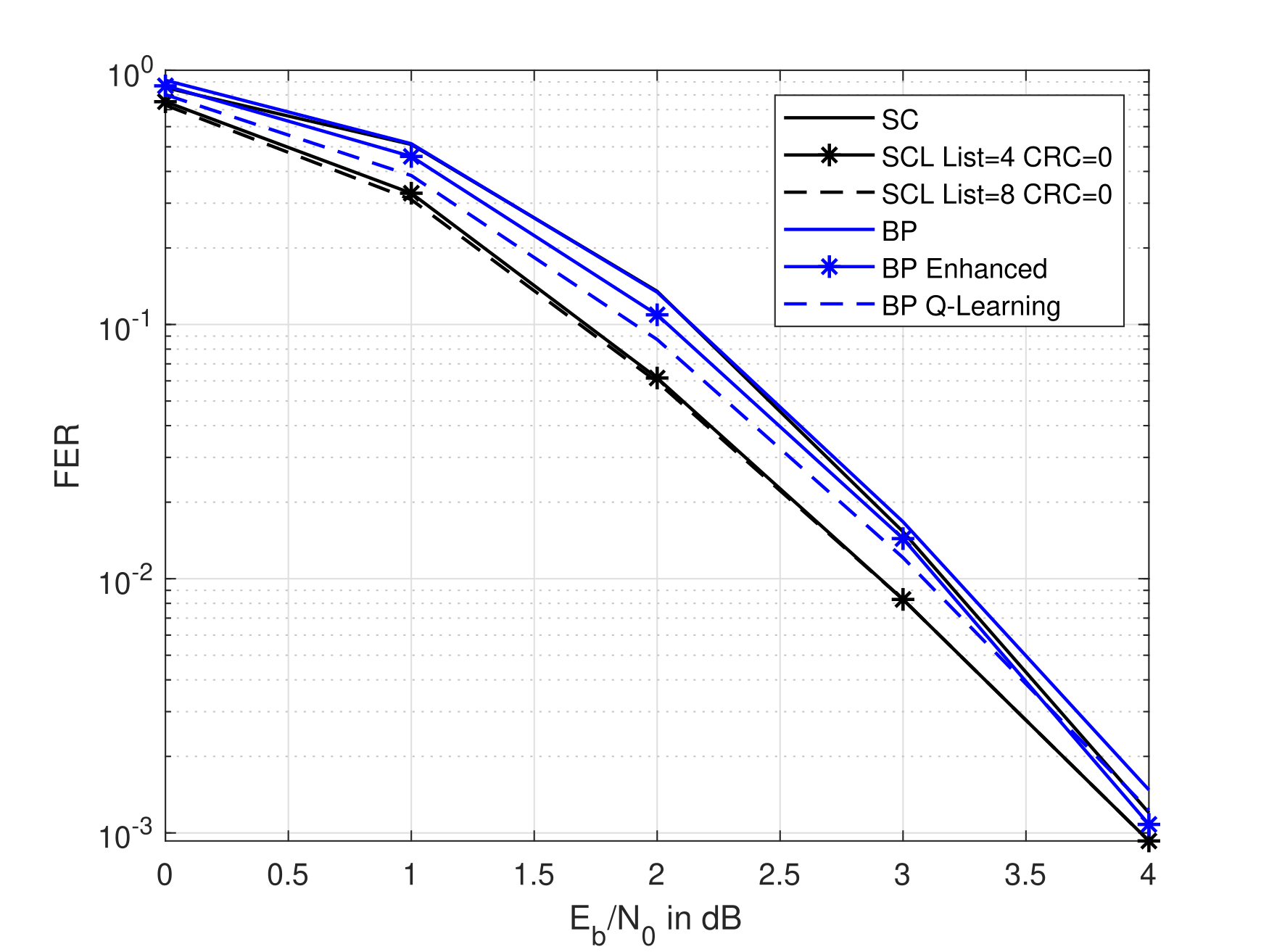} 
	        \caption{FER comparison between Arikan's original SC, SCL with List-4 and List-8, Arikan's original BP,  our proposed BP Enhanced and BP Q-learning decoders for N=256 and K=128; no CRC used.} 
	        \label{figura3} 
	        \vspace{-0.15in}
        \end{figure}

    \hspace{0.2in} As can be seen in Figs. \ref{figura2} and \ref{figura3}, for N=256, R=$\frac{1}{2}$, both results have shown that the proposed Q-Learning-Driven BP decoder outperforms the SC, BP, and Enhanced BP algorithms by up to approximately 0.5 dB and 0.4 dB in terms of BER and FER, respectively, at 2 dB, and approaches the performance of the benchmark decoder in the literature, SCL. However, note that the performance gain decreases as the signal-to-noise ratio increases. Thus, the proposed QLBP algorithm is more suitable to be implemented at lower SNRs at this code length.

\begin{figure}[h!]
		    \centering
	        \includegraphics[width=2.5in]{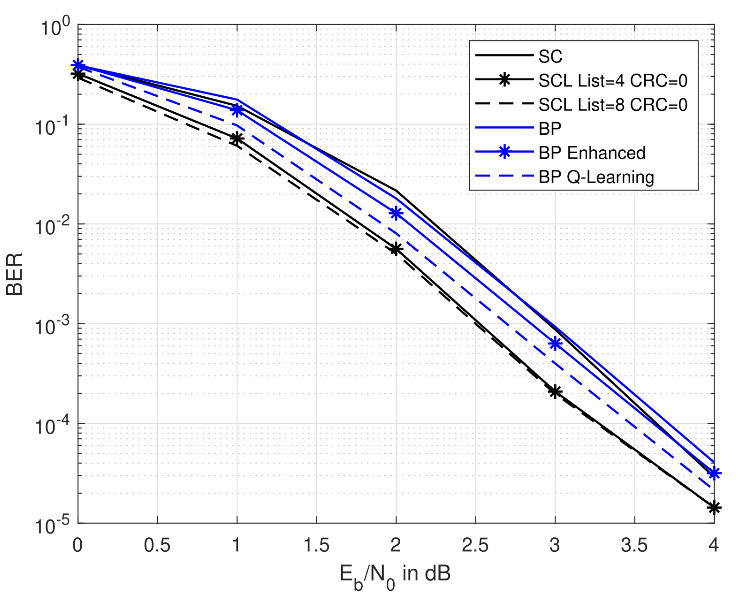} 
	        \caption{BER comparison between Arikan's original SC, SCL with List-4 and List-8, Arikan's original BP,  our proposed BP Enhanced and BP Q-learning decoders for N=512 and K=256; no CRC used.} 
	        \label{figura4} 
        \end{figure}

\begin{figure}[h!]
		    \centering
	        \includegraphics[width=2.5in]{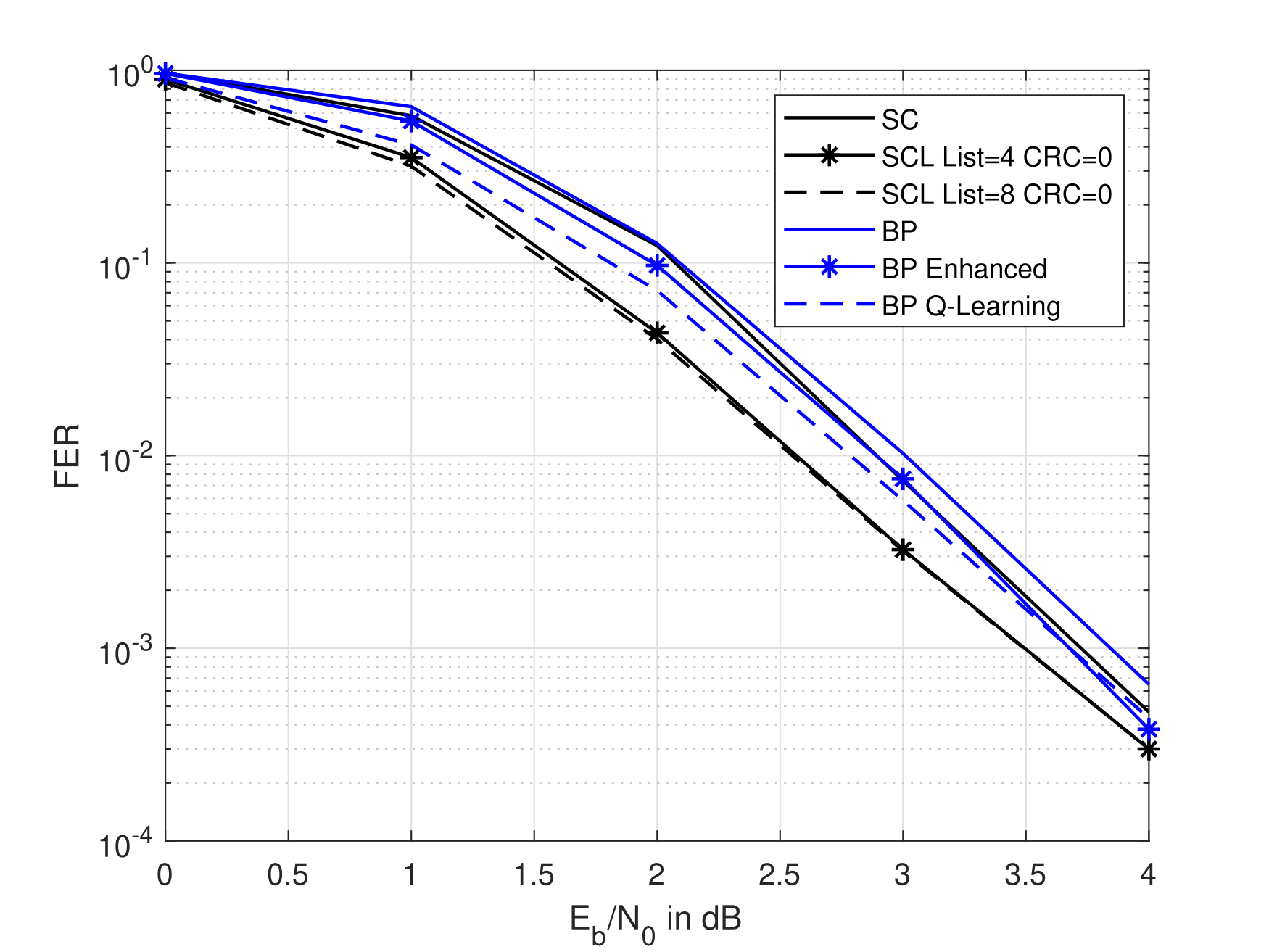} 
	        \caption{FER comparison between Arikan's original SC, SCL with List-4 and List-8, Arikan's original BP,  our proposed BP Enhanced and BP Q-learning decoders for N=512 and K=256; no CRC used.} 
	        \label{figura5} 
	        \vspace{-0.15in}
        \end{figure}

\newpage
    \hspace{0.2in} In the second example, we assess the decoders for N=512 and K=256 in Figs. \ref{figura4} and \ref{figura5}. As mentioned before, the proposed QLBP decoder also outperforms the SC, BP, and Enhanced BP algorithms by up to approximately 0.4 dB and 0.25 dB in terms of BER and FER, respectively, at 2 dB. Besides that, although the performance of the SCL decoder has not been achieved, the proposed QLBP decoder performance has got even closer to the benchmark decoder and renders itself more easily to parallel implementation. The performance gain of BER still maintained constant over the SNRs even though the FER has not.

    \hspace{0.2in} Future works will consider puncturing techniques \cite{ref11}, multiple-antenna systems \cite{ref12}, \cite{ref13}. 

\section{Conclusion}

  This paper has investigated the design of a BP decoder driven by Q-Learning, which seeks the best action, a weighting factor, for a specific state, an input in the processing element. More specifically, from our experience Q-learning learns the optimal policy that maximizes the total reward, that is, a successful decoding. Thus, in the long term, the decoder learns how to weigh each processing element. Finally, simulations have shown that the performance of the proposed  QLBP decoder for Polar Codes is better than Arikan's SC and BP codes, and closely approaches the performance of the SCL decoders.

\end{document}